# Systematic investigation of methods for multiple freeform optimization in multi-lens imaging systems

**CHANG LIU,[1,*] HERBERT GROSS,[1]**

[1]*Friedrich-Schiller-University Jena, 07743 Jena, Germany*
**Corresponding author: chang.liu@uni-jena.de*

**With the development in freeform technology, it has now become more and more feasible to use freeform surfaces in real system designs. While the freeform surfaces helping optical designers achieve more and more challenging system features, the methods for multiple freeform implementations are still underdeveloped. We therefore investigate strategies to use freeform surfaces properly in imaging optical systems with one Scheimpflug system and one lithographic system. Based on the studies of the influences of the freeform normalization radius, freeform order and system eccentricity, the methods of determining the optimal location for implementing one freeform surface are discussed. Different optimization strategies to optimize two freeform surfaces are discussed to compare their resulting influences on the system performance. On top of that, ways to implement more than one freeform surface in the optical system is also investigated. In the end, a workflow is presented as guidance for implementing multiple freeform surfaces with respect to system aberration constitutions.** 

## 1. INTRODUCTION

The development in optical manufacture, alignment and testing [1,2] has enabled the increasing use of freeform surfaces in many kinds of optical systems [3,4]. The demanding system requirements need the involvement of optical surfaces that are able to provide more degrees of freedom. For better and more efficient use of the freeform surfaces, the understanding of freeform surfaces from different perspectives is necessary[5,6,7,8]. One important aspect for optical designers is thereby the question of where to put the freeform surfaces to make the best use of them, and how to optimize the system afterwards. These questions are rarely discussed in the publications until now. Yabe proposed a method to determine the optimal positions of aspheric surfaces continuously in systems in 2005 [9], and later extended it by adding surface tilts [10]. The extended method is able to help break the rotational symmetry of the initial system, creating axially asymmetric optical systems. The tilted aspheres, though can be treated as freeforms, are still different from the freeform surfaces we normally refer to. They have fewer parameters and fewer degrees of freedom, and the decoupling of sagittal and tangential planes are also not possible. With the development in freeform mathematical representation, fabrication and mounting, it is now more feasible to use freeform surfaces that provide more correction abilities to help achieve better system performance.

The term freeform in general refers to surfaces with arbitrary shapes. In the context of optical design, freeform surfaces are defined as surfaces that are not spherical anymore. We further narrow the definition down to use non-rotationally symmetric surfaces as our choice of the freeform. The reason for doing this is the development on rotationally symmetric aspheres, which can also be treated as freeform surfaces, has made the traditional asphere well accepted by the optical industry. The non-rotationally symmetric freeforms, on the other hand, has been treated with a more conservative attitude due to the difficulties in improving fabrication and alignment precision. The freeform surfaces have to be used in the most efficient way to benefit the system performance.

One of the primary concerns for feasible freeform optical designs is the number of freeform surfaces needed to achieve the required specifications. Considering the same tolerancing condition, the goal is hereby the minimization of the required number of freeforms. The location of the freeform has to be chosen in a way that aberrations are compensated as good as possible. For challenging systems, one freeform might sometimes not be sufficient. In that case, one needs to consider the question of where to put the second freeform, and how to optimize both freeform surfaces appropriately. For reflective systems like the TMA [11] with only a small number of surfaces, these questions might be less relevant. However, when complex refractive systems containing more optical components are considered, questions of placement and optimization of the freeform surfaces become increasingly important.

In the content of this paper, we conduct a numerical case study for multi-lens systems. We analyze two challenging systems, the Scheimpflug system and the lithographic system, regarding their improvability with freeform surfaces. The Scheimpflug system has both large spherical aberration and large field related aberrations, while the lithographic system has dominating field aberrations. The empirical insights obtained from this work are suitable for the investigated systems, and can also provide some insights regarding the use of multiple freeform surfaces in general.

In this paper, we first introduce these two investigated systems in section 2. The precautions for using freeform surfaces are discussed in section 3. The strategies to place freeform surfaces in a system when more of them are needed are discussed in section 4. The two-freeform position issues are also investigated in section 4. We provide in section 5 a workflow for using freeform surfaces in system optimization. The application and advantages of the given workflow are also presented in section 5.

## 2. Investigated systems and software

We choose two imaging systems, a Scheimpflug system and a lithographic system as indicated in Fig. 1 and Fig. 2, to investigate. Both systems are very challenging for conventional optical design with solely rotationally symmetric surfaces.

The Scheimpflug system's object plane is tilted by 70° and has an averaged working f number of 1.08 for the image side. The system has only spherical surfaces at this point. However due to the tilt of the object plane, the system's rotational symmetry is broken. A good correction of the introduced large field aberrations is hard to realize with only rotationally symmetric surfaces. This makes the Scheimpflug system a great candidate for implementing freeform surfaces. The dominating aberration is spherical aberration, along with large contributing coma and astigmatism. The Scheimpflug system has a large inherited quasi-paraxial keystone distortion that cannot be corrected, and therefore the main concern of the merit function formulated for this system is the rms spot radius error. In the original merit function, the average distortion correction contribution is only 1%, in contrast to the nearly 20% average contribution of the resolution correction.

The lithographic system [12] on the other hand has a decentered object field to avoid the central obscuration introduced by the mirror components. The rotational symmetry is already broken when the mirrors are used. The numbering of the surface starts from 2 because the first surface is a coordinate break surface for system decenteration. Many lenses are used for this system to reach an image side working f number of 0.43, which makes it even more difficult to decide where to place the freeform surfaces. There are also two intermediate image positions in the system, which gives us an extra access to the system's field plane. The original patent [12] uses twelve aspheres, while the new system is re-optimized with only spherical surfaces, and leaves plenty of room for improvement. Field aberrations are dominating in the case of the lithographic system due to the laterally shifted field of view. The merit function of the system controls system parameters, telecentricity, distortion and resolution. Weightings of different operands are balanced so that no strong contribution occurs. For example, the averaged contribution coming from the telecentricity correction is 5% for the start system. Rougly 9% contribution in average comes from the resolution correction over five discrete fields in y direction, and 2% in average comes from the distortion correction operand. The merit function of both systems stay unchanged during our investigation, but the contribution will change automatically during optimization. The numbers are provided here to present a general description of the constitution of the merit function.

The results of the optimization are strongly related to the constitution of the merit function. The optimal locations for freeform surfaces are also somewhat changing with different merit functions. Therefore the formation of the merit function is not trivial. It should be well balanced and chosen correctly to meet the requirements of system specifications.

All optimizations and investigations are done with the optical design software Zemax 13. As we already know, the optimization is nothing but the search for global minimum in the solution space. With the same constraints, the more degrees of freedom of the system parameters are allowed, the more complicated the solution space will be, but also the more likely a smaller minimum can be found. Even for the same start system, due to the complicated solution space, similar optimizations can lead to different solutions. This can be problematic when we are trying to conduct a systematic numerical investigation comparing the results of different optimizations. What we have noticed during our work is, though the results of similar optimization actions may not be the same, the values of the merit function are still in a comparable range. We just need to keep the uncertainties of the optimization algorithm run in mind when we analyze the obtained results.

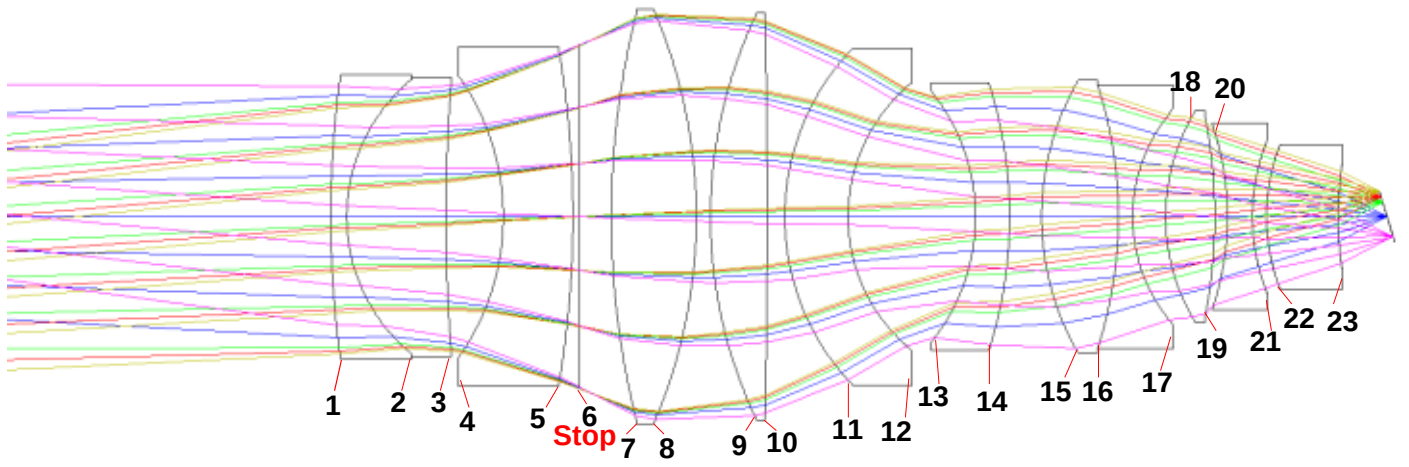

Fig. 1 Layout of the Scheimpflug system, with the stop placed at surface 6. The object plane is tilted towards the lenses by 70° and not shown in the layout.

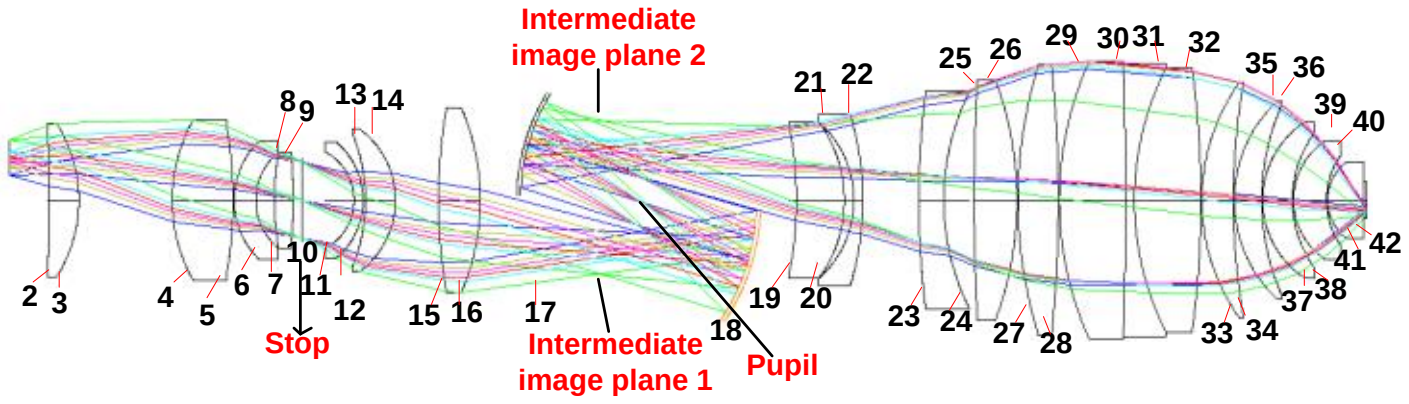

Fig. 2 Layout of the lithographic system. The system stop is surface 10. The system also has two intermediate image planes between surface 17 and surface 18.

## 3. System optimization with one freeform surface

To simplify our discussion, we start with the simpler case: optical system optimization with one freeform surface. We use the freeform surface type 'Zernike fringe sag'. Different freeform surface types have different coordinate systems and orthogonality. Based on our previous experience [6], freeform surface types with slope and spatial orthogonality show no significant differences regarding the optimization performance for selected system types. We therefore choose a freeform surface type with spatial orthogonality, which is already implemented in the software.

Many aspects need to be considered while using freeform surfaces in Zemax. Factors like the size of the area the freeform is defined on, the number of orders of the freeform, or the optimal freeform position with respect to the stop position are not trivial. We are going to discuss these topics in the context of this section.

**A. Impact of the normalization radius**

The freeform terms are not defined directly on the clear aperture; instead, a normalization radius is defined on which the high order terms are valid. If we assume the ratio between the normalization radius and the clear aperture to be k,

$$k = \frac{norm._radius}{clear_aperture}$$

the chosen value of k will have an impact on the aberration contribution of the freeform.

When k<1, which means the normalization radius is smaller than the surface diameter, the freeform is defined on a smaller area. If one does not allow extrapolation for the area outside the normalization radius, like the case indicated in Fig. 3, a step-like discontinuity will appear. The surface area outside the defined freeform stays spherical, therefore no high order correction is possible for the outer ring rays. Extrapolation should be used in any case for a proper correction of the complete ray set. If one allows for extrapolation, the outer rays will be considered. But the edge of the freeform surface cannot be controlled properly, and it can appear to be quite unreasonable as shown in Fig. 4.

When k>1, which means the normalization radius is larger than the surface diameter, only the central area of the defined freeform is actually used in the system. When the normalization radius is much larger than the surface diameter, the central part is relatively flatter than the edge in the case of lower orders. Consequently, the ability of the freeform surface for correction is weaker. One needs to bring in more oscillation in the central part by using higher orders.

To study the choice of the optimal one-freeform surface location, we implement a workflow in a Zemax macro. This macro is able to plot and compare the resulting performance as consequences of putting freeforms at different surface locations. Information of the system symmetry is input into the macro to define the distribution of freeform variables. For example, for x symmetric systems, only x symmetric Zernike terms are used as variables, while the rest stay unused. The reason for that is, only x symmetric terms are contributing for the correction of x symmetric systems. However, Zemax is not able to detect system symmetries and will use all preset variables for optimization. The large number of variables does not only increase the complexity of the solution space, but also decrease the optimization speed. Additionally, we only use the variables of the freeform surfaces and the imaging distance of the system for optimization for a fixed number of cycles. The reasoning for that is not only the improvement of optimization convergence, but also to avoid the influences of other surfaces. The general rule for placing one freeform is better seen when the freeform is studied independently. A plot indicating the resulting performance of placing freeforms at different surface locations is given after running the macro, with which, the decision of where to place one freeform surface becomes clearer.

We run the macro on the Scheimpflug system with different normalization radii. Fig. 5 shows the impact of different k values on the system with decreasing freeform orders. The x axis represents the surface number where freeform locates, and the y axis represents the merit function value after optimization. Each dot on the figure represents the performance of the system after making the corresponding surface freeform. In this case and most of the following cases, we use the merit function value as a comprehensive criterion for system performance evaluation. Four k values ranging from 0.5 to 4 are chosen allowing for extrapolation. Fig. 5(a) (b) (c) have the same y axis scale for direct comparison. We can see from the case in Fig. 5(a) that, with the highest freeform order, the differences between different k curves are the smallest. The differences increase as the order decreases. This agrees with the explanations discussed above.

To guarantee the best performance of the freeform surface, the k has to be set not far away from the clear aperture. What cannot be seen from Fig. 5 is the freeform surface sag value, considering the possible edge problem when k is smaller than 1, we set the k to be 1.5 for our following studies.

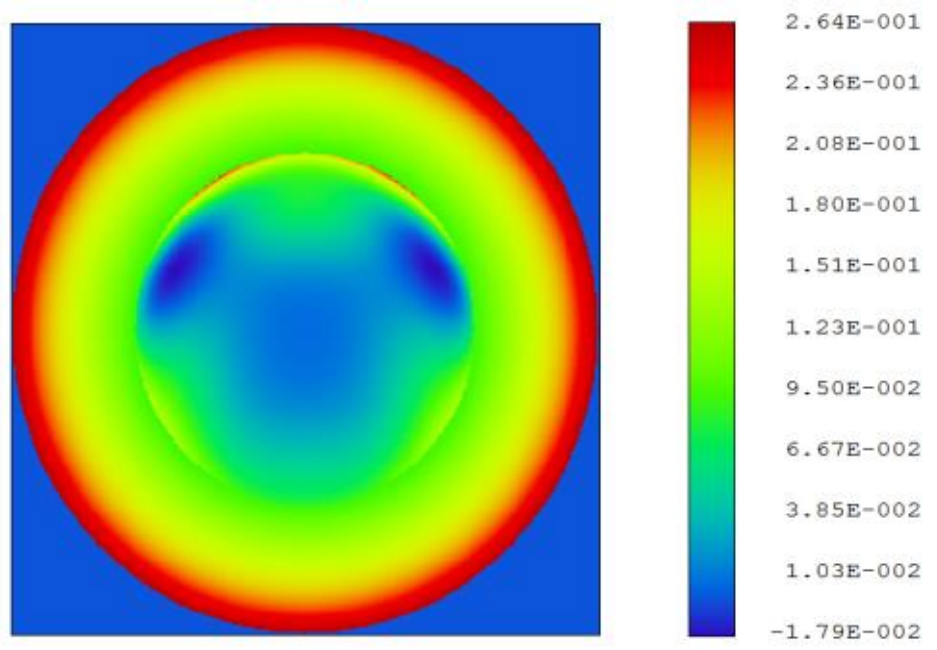

Fig. 3 Freeform surface sag, k<1, no extrapolation. The freeform surface sag is not continuous. The edge sag is 0.26mm.

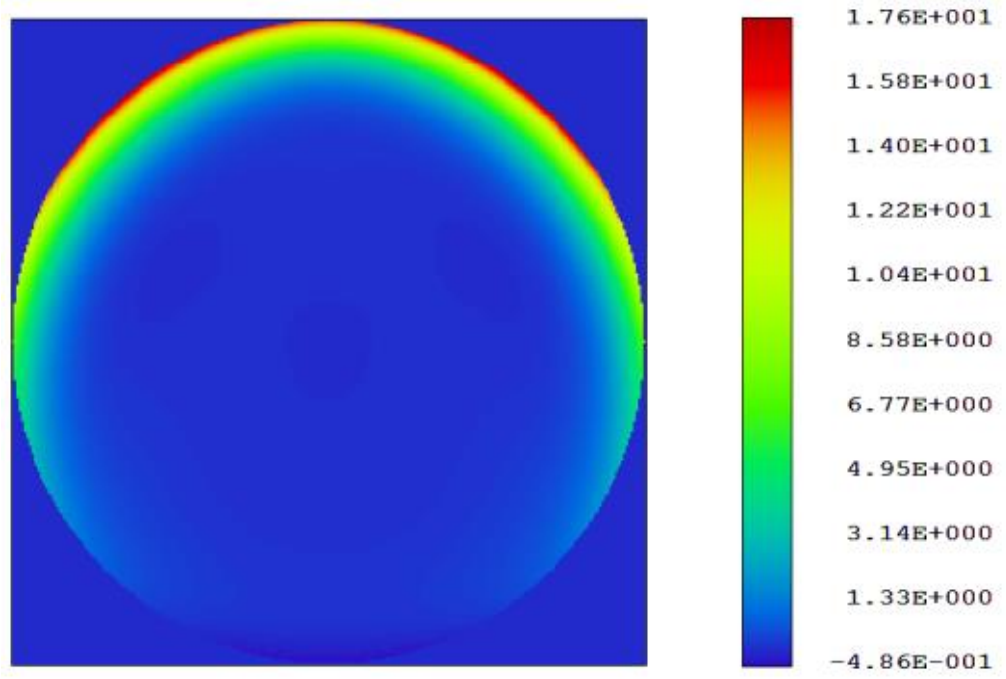

Fig. 4 Freeform surface sag, k<1, with extrapolation. The freeform surface has a large edge sag of 17.6 mm.

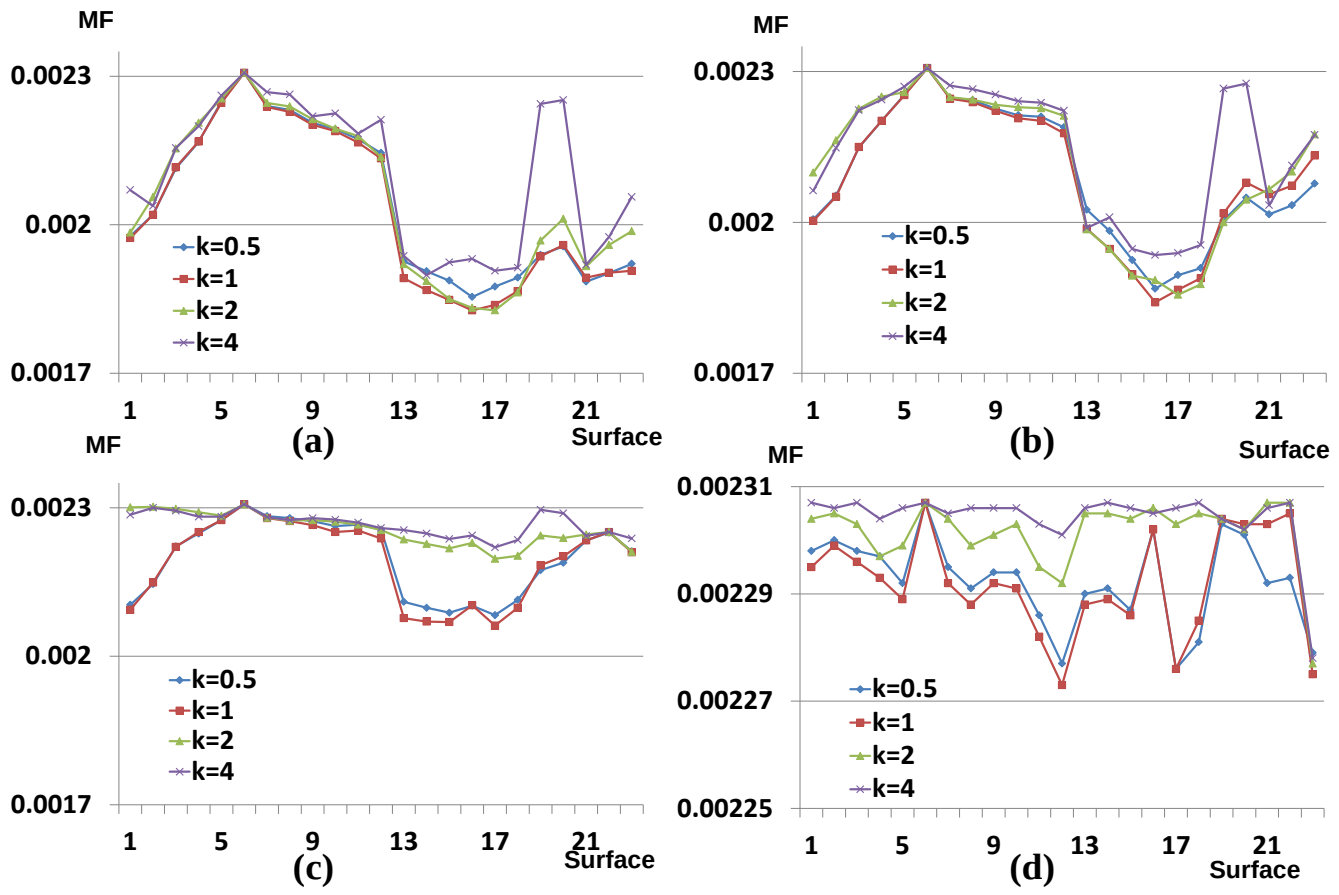

Fig. 5 Scheimpflug system, impact of the normalization radius with the decreasing order. (a) freeform order up to the 10th, (b) freeform order up to the 8th, (c) freeform order up to the 6th, (d) freeform order up to the 4th. The y axis corresponds to merit function value, the x axis corresponds to freeform surface position.

### B. Eccentricity

Unlike spheres and aspheres, the break of rotational symmetry of the freeform surfaces makes them especially beneficial for correcting off-axis aberrations, which exist due to the asymmetry of the off-axis ray bundles. Intuitively, the freeform surfaces work better at locations where ray bundles from different fields are more separately. To support this claim, a parameter called eccentricity [13] should be introduced that describe the relative position of the surface with respect to stop and field locations.

$$\chi = \frac{h_{CR} - h_{MR}}{h_{CR} + h_{MR}}$$

In this equation, $h_{CR}$ represents absolute chief ray height, $h_{MR}$ represents absolute marginal ray height. The value of $\chi$ equals to 1 for the image plane where ray bundles are completely separated, and equals to -1 for the pupil plane where all ray bundles intercept. Therefore, the normalized eccentricity $\chi$ is able to describe the separation between the selected off-axis bundles and the reference on-axis field. For the two investigated systems, only fields in y direction are of interest due to the broken symmetry in this direction. Therefore, the chief ray heights and marginal ray heights in y are used for the calculation of eccentricity.

Now that we have set up the parameter of eccentricity, we are going to make each non-planar effective surface of our investigated systems a freeform successively using the macro.

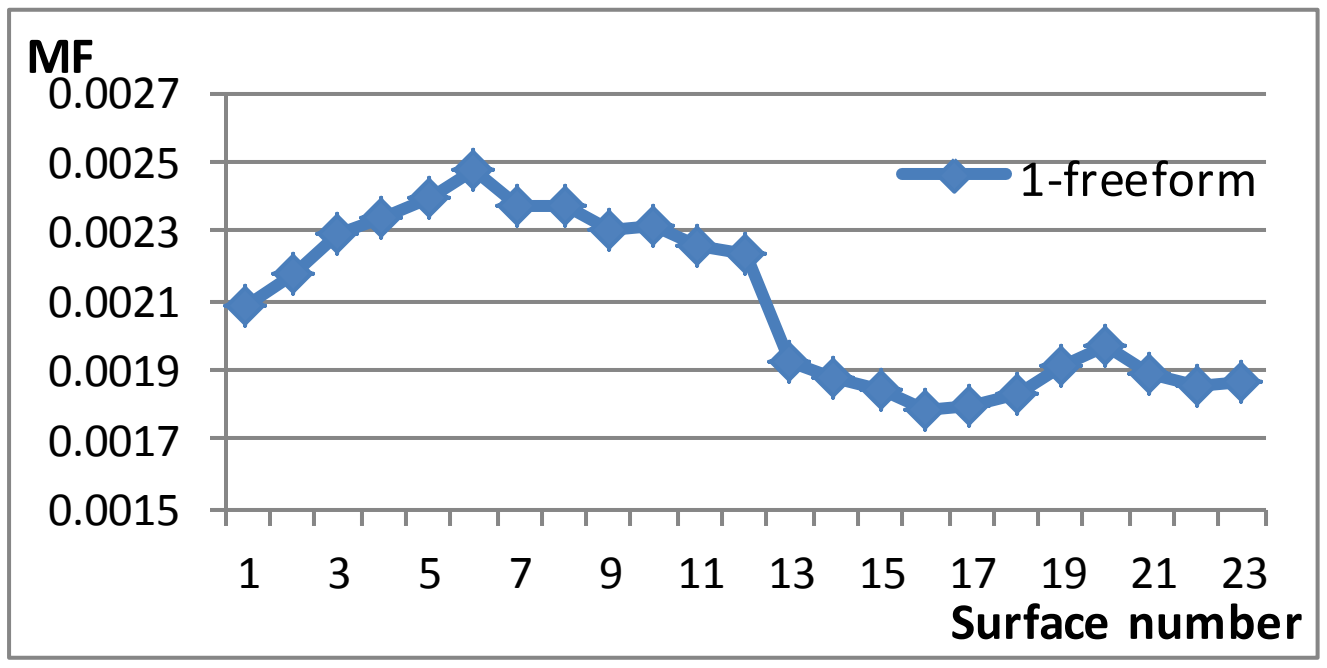

(a)

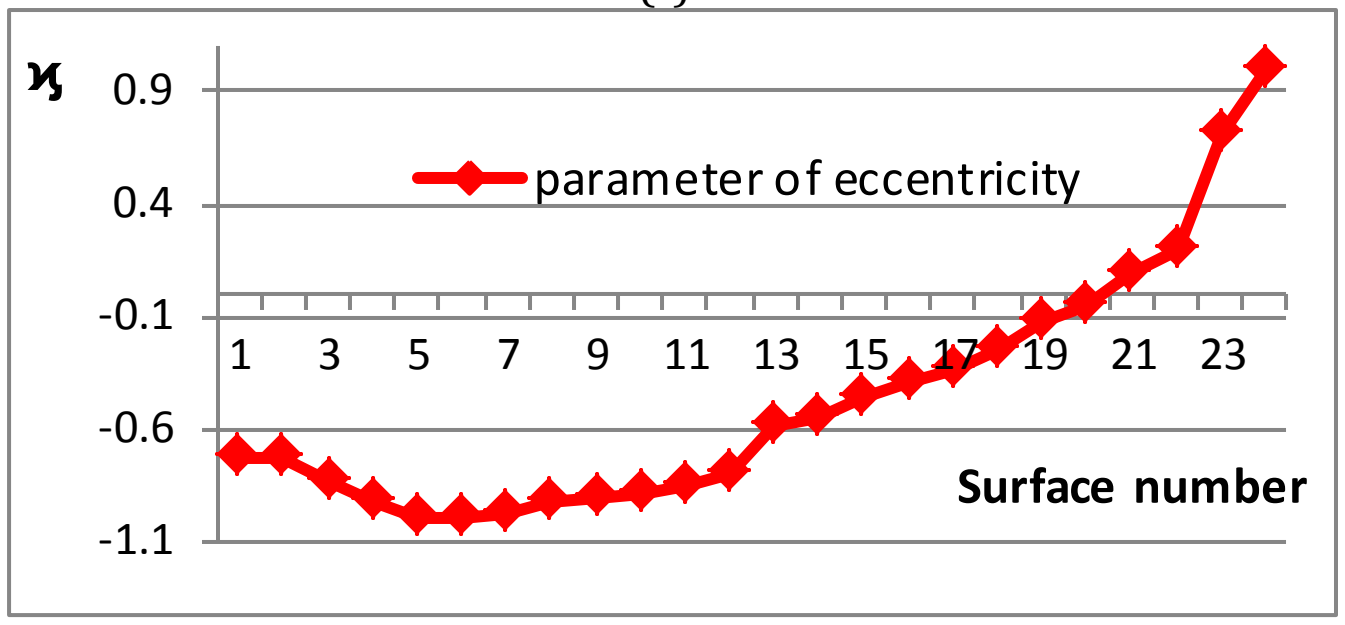

(b)

Fig. 6 Scheimpflug system, one freeform performance vs. eccentricity. The x axis represents the freeform surface number, and the y axis represents the absolute value of the merit function for the one-freeform curve (a) and eccentricity for the eccentricity curve (b).

In Fig. 6, the one-freeform performance of the example Scheimpflug system is shown together with the eccentricity curve. The system stop is a dummy surface placed at surface 6. The diagram's y axis stands for the value of the merit function for the one-freeform curve, and the value of eccentricity for the eccentricity curves. As we can see from the figure, there are certainly some similarities between the freeform curve and the eccentricity curve. Freeform positions close to the stop, which correspond to the small value of the eccentricity curve, have worse performance than the freeform positions away from the stop.

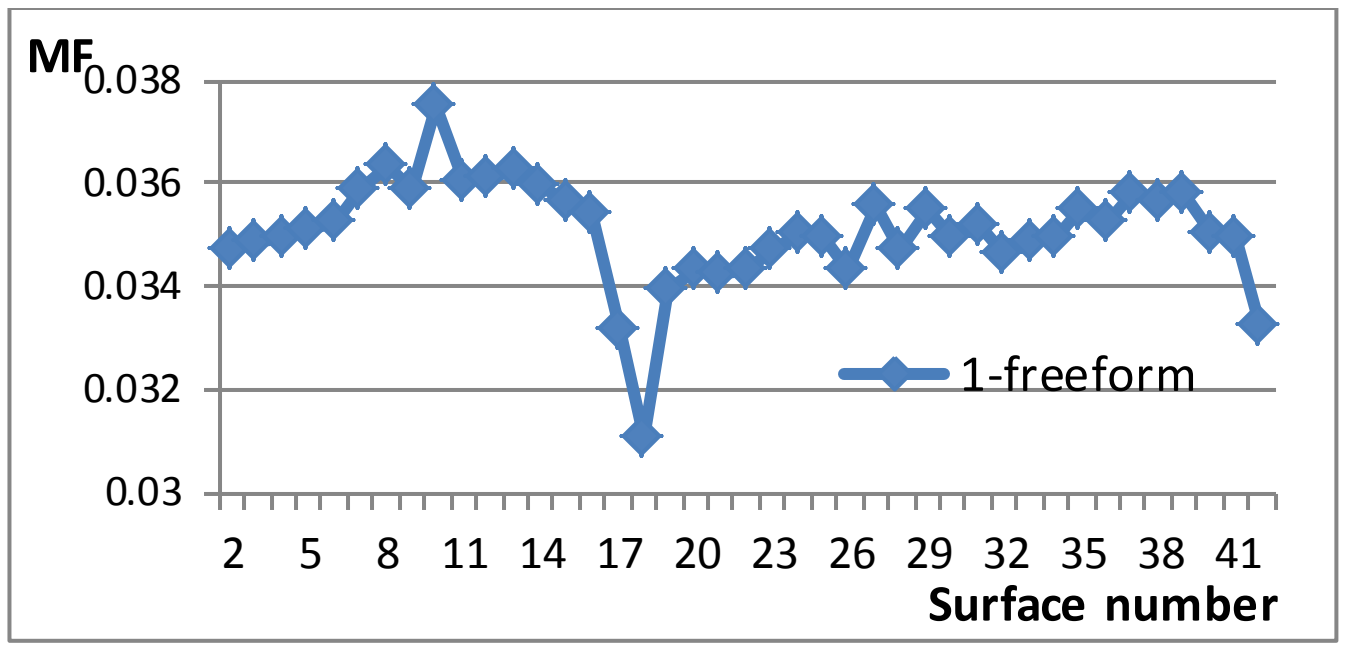

(a)

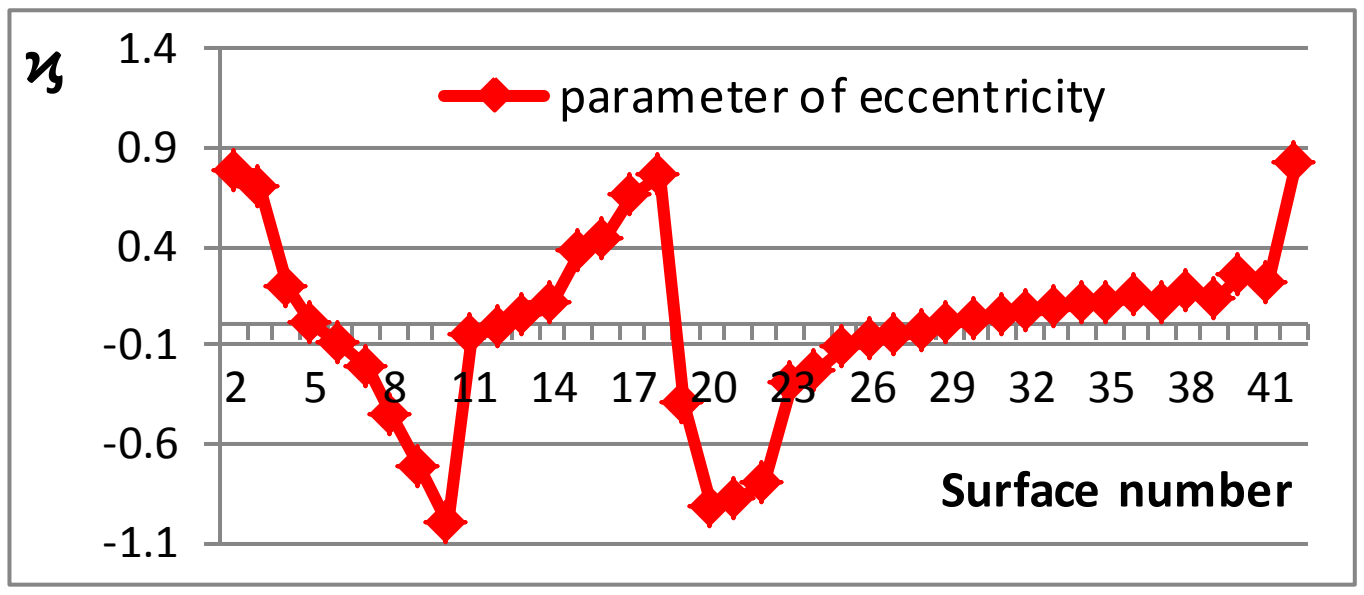

(b)

Fig. 7 Lithographic system, one freeform performance vs. eccentricity. The x axis represents the freeform surface number, and the y axis represents the value of the merit function for the one-freeform curve (a), and eccentricity for the eccentricity curve (b).

Another example is given in Fig. 7 for the lithographic system. The lithographic system has two intermediate image planes between 17 and 18, and the stop is at surface 10. The peaks and dips at surface 10 and surface 18 of the eccentricity curve somewhat correspond to the dips and peaks of the one-freeform position curve, and the pupil position is again the worse choice for placing the freeform surface.

Both selected systems have large field-dependent aberrations, especially the lithographic system. Therefore, the correction of these aberrations is more challenging and works better at locations away from the stop with larger chief ray heights. For the Scheimpflug system, correction of the spherical aberration is also necessary, which works more beneficial at locations near the stop. The trade-off between these two could explain the turning of the one-freeform curve in the rear part. With the knowledge of the similarities between the one-freeform curve and the eccentricity curve, we can evaluate the positions for placing freeform surfaces very fast without running the macro. This can provide fast assessment of the system as a guide. Nevertheless, the detailed freeform position information can only be obtained via running the macro. With only the freeform surface and the imaging distance as variables, the speed of the macro has improved drastically.

### C. Impact of the freeform order

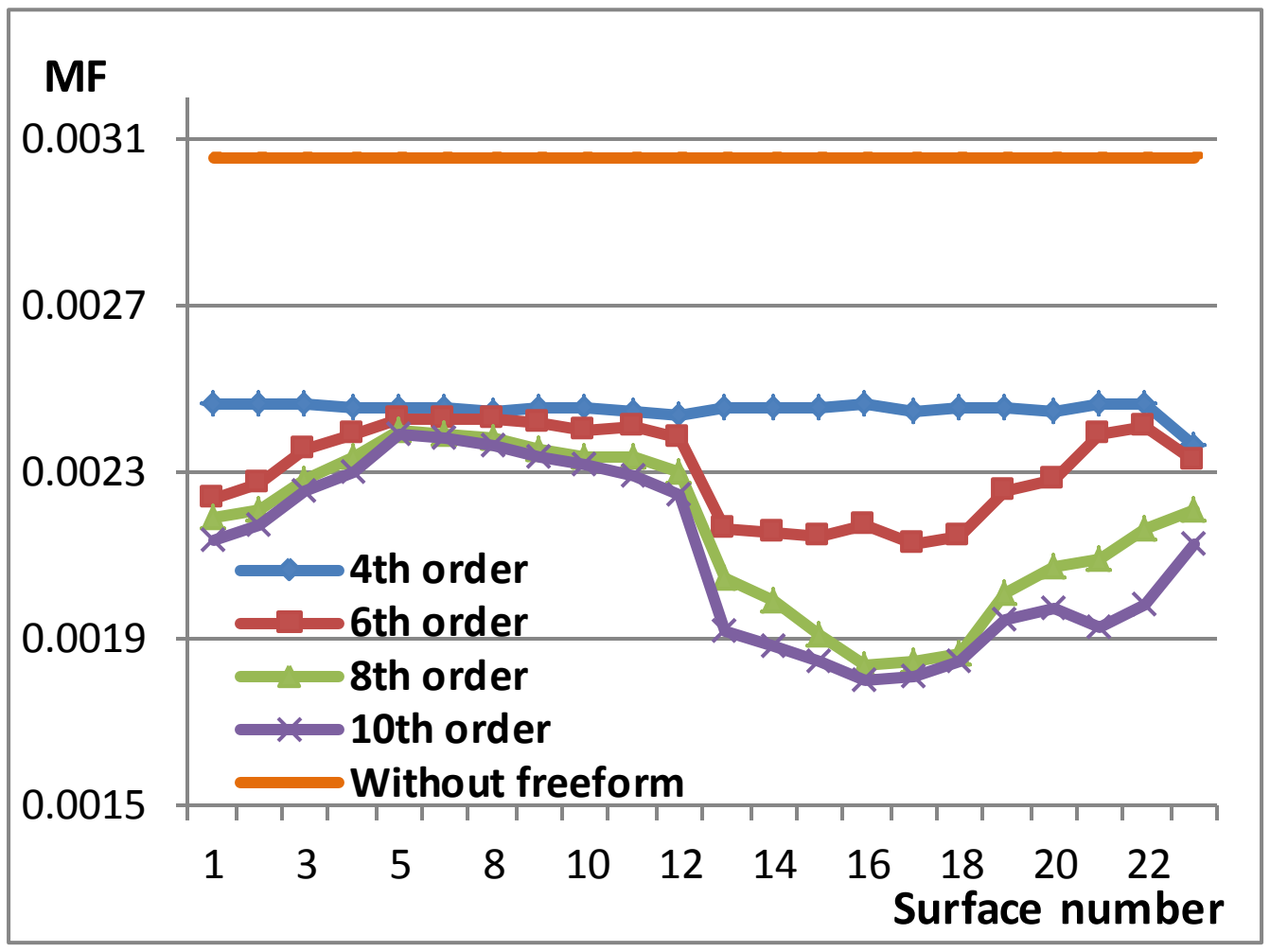

Fig. 8 Scheimpflug system, impact of the freeform order. The y axis represents the merit function value, the x axis represents the freeform surface number. Colored curves correspond to one-freeform performance using freeforms up to the respective orders. The merit function value without freeform is also given for comparison, the performance improvement factor is roughly 1.7 for the optimal surface.

Up until this point, we have used the complete program-provided set of the freeform terms up to the 10th order according to the ordering of the Zernike fringe surface. The order is easily expandable by using a user defined Zernike surface type. But the question arises if it really necessary to use that many high orders for every system. For the optimum correction results, as well as the fast first-hand evaluation, we should optimize the number of orders included in the macro for systems with different aberration constitutions.

It is well known that, with higher orders, higher spatial frequencies of the surface sag are introduced, which can potentially make the production of the surface more difficult and even unrealizable. For a good design, the qualified system performance should be achieved with the simplest freeform surface sag, which in most cases means the smallest freeform order. On the other hand, one of the main advantages of freeform surface is its ability to correct high order aberrations with its high order terms. For systems whose high order aberrations are enormous, or even dominating, freeform surfaces of at least the equal order are necessary for a good correction.

When more high order terms are set as variables in the macro, the complicated solution space requires more running time for each surface evaluation. Moreover, during the process of increasing freeform orders, the contribution coming from the highest orders is getting smaller. Summarizing all the above points, it is therefore neither necessary nor convenient to include too many high order terms for the freeform surfaces.

We want to consider again the Scheimpflug system as an example. The system's main aberration contribution includes Zernike aberrations Z11, Z12, Z15 and Z21, which belong to the 6th and the 8th order. Of these aberrations, Z21 has the smallest contribution. As we can see from Fig. 8, the system performance is greatly improved with freeforms, while the curve with only freeforms up to the 4th order is significantly worse than the other curves. The system performance with freeforms up to the 6th order is already quite good, considering most of the main aberrations are of the 6th order. When we further increase the freeform orders, the higher order aberration Z21 can also be well compensated, which leads to the further decrease of merit function values. At this stage, we can conclude the freeform surface now contain enough orders for a correct selection of the position. Further increase of the order will not help tremendously with the system performance, and is not necessary for a correct run of the macro, where we merely want to evaluate the location for the best one-freeform surface. For the Scheimpflug system, the best one-freeform surface location is surface 16.

For the lithographic system, the main aberration contributions are Z5 and Z11. Z9 and Z17 are also very large, followed by Z12, Z20 and Z27, which are significantly smaller. The impact of the freeform orders for the system performance is shown in Fig. 9. Compared with the original system performance without any freeforms, the optimized system is in the best case better. The large aberration Z11 cannot be well corrected by a 4th order freeform, and the blue curve has a worse performance than the rest. While 6th order aberrations are compensated by the 6th order freeform surfaces, the remaining 8th aberration Z17 needs a higher order. For fast evaluation and determination of the optimal freeform location, freeform surfaces up to the 8th order should already be enough.

We want to note that, knowledge of the aberration theory up to the 8th order is necessary here in order to solve the considered problems analytically, but the current available formulas for nodal aberration theory [14,15] suitable for freeforms only include orders up to the 6th for the basic shape. It is also extremely complicated to consider the overall resulting effects of every added high order terms[16]. Therefore the analytic solution is neither realistic nor substantial momentarily for solving freeform positioning problems for complex multi-lens systems. But surface-resolved contribution representation for higher orders [7] helps to understand the compensation structure of a system, which usually indicates appropriate strategies for correction. Due to the re-distribution of aberrations, numerical procedure is necessary.

With the correct selection of the freeform orders used for freeform position assessment, we are able to further increase the macro speed, and find the optimal surface position within the shortest possible amount of time without losing accuracy.

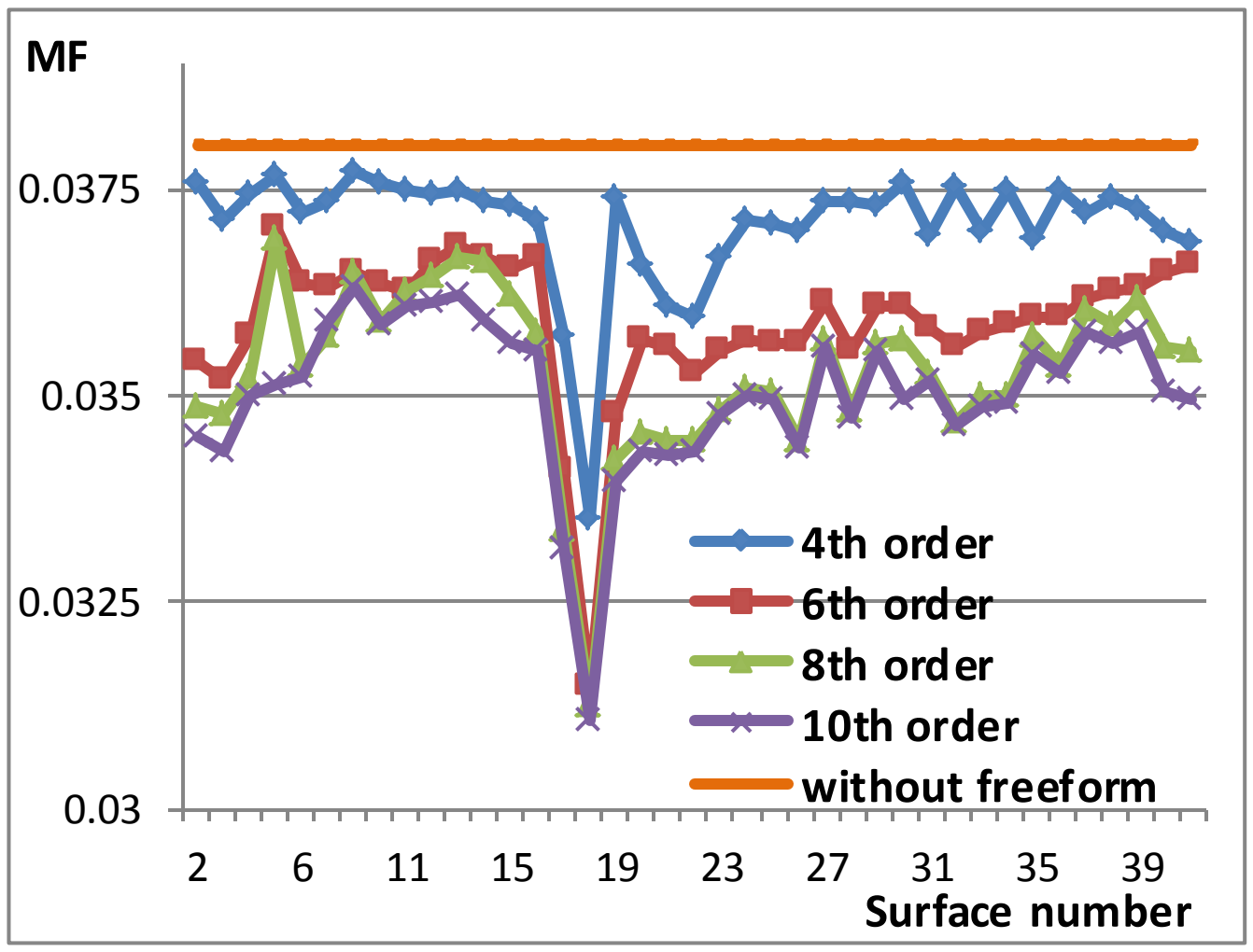

Fig. 9 Lithographic system, impact of the freeform order. The y axis respresents the merit function value, and the x axis represents the freeform surface number. The colored curves correspond to one-freeform performance using freeforms up to different orders. The merit function value without freeform is also given for comparison, the performance improvement factor is roughly 13.

## 4. System optimization with two freeform surfaces

For many systems, the design strategy we discovered in section 3 to implement one freeform is already enough to reach the required system performance. Nevertheless, complicated systems sometimes need more than one freeform. If more freeforms are introduced to the system, it is not only more difficult to determine the freeform locations due to the increased possibilities, the potential interactions between multiple freeforms are also worth studying. Considering the aberration theory, it could be beneficial to select one freeform at locations with $\chi$ close to -1 for field-independent resolution improvement, and select the other freeform at locations with $\chi$ close to 1 for field-dependent aberration correction. Nevertheless, the exact selection of freeform positions still dependent on the system aberration constitution.

### A. Optimization strategy

One important aspect for optical designers when using more than one freeform surface in an optical system is the considered optimization strategy. Thereby the questions arise if the freeforms should be optimized successively or simultaneously, and if the orders of the freeforms should increase successively or simultaneously. Due to the complexity of the investigations, we only discuss the case of two freeform surfaces in one single system. The conclusion here should be applicable for cases with more freeform surfaces.

Based on personal preferences of optical designers and correction principles of freeform surfaces, we investigate the following possible strategies that can be applied when using two-freeform surface optimization.

A. Two surfaces are optimized successively with fixed orders.
B. Two surfaces are optimized simultaneously with fixed orders.
C. Like strategy B, but the optimization cycles are done in smaller successive steps.
D. Two surfaces are optimized simultaneously with orders increased step by step.
E. Two surfaces are optimized successively with orders increased step by step.

For all strategies, only variables of two freeform surfaces, as well as the system imaging distance are used for optimization to eliminate the interaction of other surfaces. Strategy A and B are designed to investigate the impact of optimization sequence. For strategy A, we add the first freeform surface, optimize for 50 cycles, and then add the second freeform surface, optimize both surfaces together for another 75 cycles. While for strategy B, two freeform surfaces are added and optimized together for 100 cycles. For systems with limited variables and not extremely complicated layout, merit functions converge quite fast, therefore 50 cycles are already more than enough for strategy A and B.

The optimization process is a nonlinear iterative process, which means the next iteration gives a better result. Strategy C is proposed because some of our colleagues experienced the benefit of dividing large number of cycles into several smaller steps. We want to verify this claim systematically. This strategy divides 100 optimization cycles into five successive steps that each contains 20 cycles. It has the identical condition as strategy B, and the results should be comparable.

Strategy D and E are suggested to figure out the better way to add freeform orders. Strategy D optimizes two freeforms together starting from the lowest order. The orders are increased step by step for both surfaces simultaneously in the following steps. Each step contains 25 cycles. In strategy E, we start the optimization with one freeform with the lowest order. This freeform will be fixed when we optimize the other freeform with the lowest order. The second freeform will again be fixed when we turn to the first freeform with a higher order. Simply speaking, both freeforms are never optimized at the same time, and the introduction of higher orders is always done in turns. The number of each cycle is 25 so that the overall cycles are still comparable with strategy A and B.

Strategy A and E are both optimizing two freeforms successively. The difference between them is, the freeform orders are increased to the maximum step by step for E, while staying fixed from the beginning at the 10th order for A.

We run different optimization strategies on the Scheimpflug system. The first freeform surface is already fixed as surface 16, but the second freeform surface is selected from the remaining surfaces. The system performance represented by the merit function value for different surface combinations and different optimization strategies are shown in Fig. 10. We can see clearly from Fig. 10 that there are no obvious differences between strategy A, B, C and D, if we consider the results of the optimization. But strategy E, on the other side, performs much worse. As can be seen, the jumping between two freeform surfaces makes the optimization more likely to be trapped in local minima with larger merit function values. The same is also true for the lithographic system. This system has a more complicated structure; therefore the uncertainty of the optimization results is also larger. The final results in Fig. 11 again show the disadvantage of strategy E in comparison with the other four. For the lithographic system, we do see slight advantages of strategy C in comparison with strategy B.

We therefore recommend implementing freeform surfaces one after another according to strategy A, which is also in correspondence with the usual habits of many optical designers. The optimization can be done in smaller steps instead of using too many cycles in one step. The systems can also be examined more frequently by using smaller optimization steps. By this implementation method, the least amount of possible freeform surfaces is needed for the investigated systems.

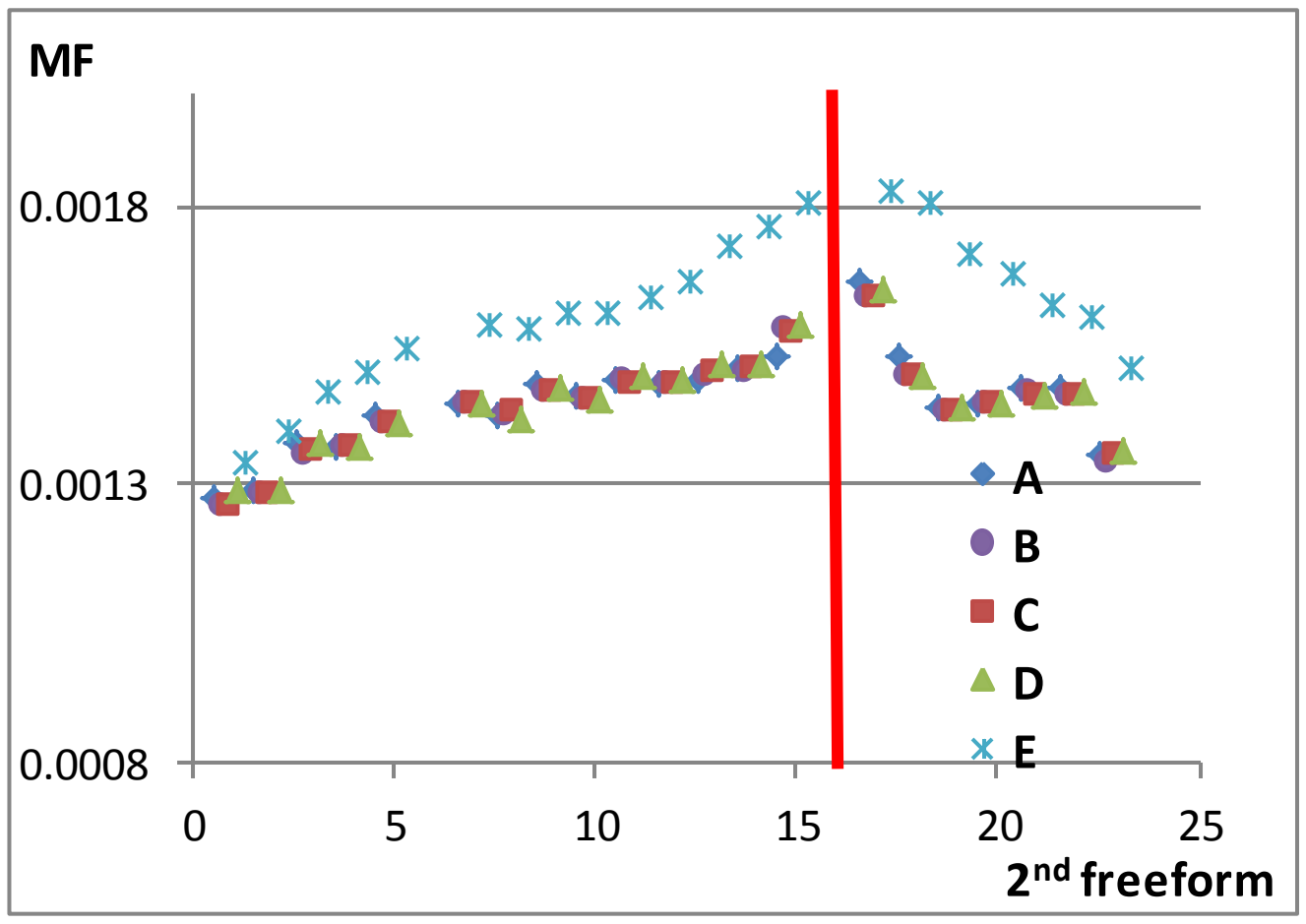

Fig. 10 Optimization strategy for the Scheimpflug system. The Scheimpflug system is selected with the first freeform surface being surface 16. The positon of the first freeform surface is marked with a red line. The y axis is the value of the merit function as the criteria for performance comparison. The x axis represents the surface number of the second freeform surface.

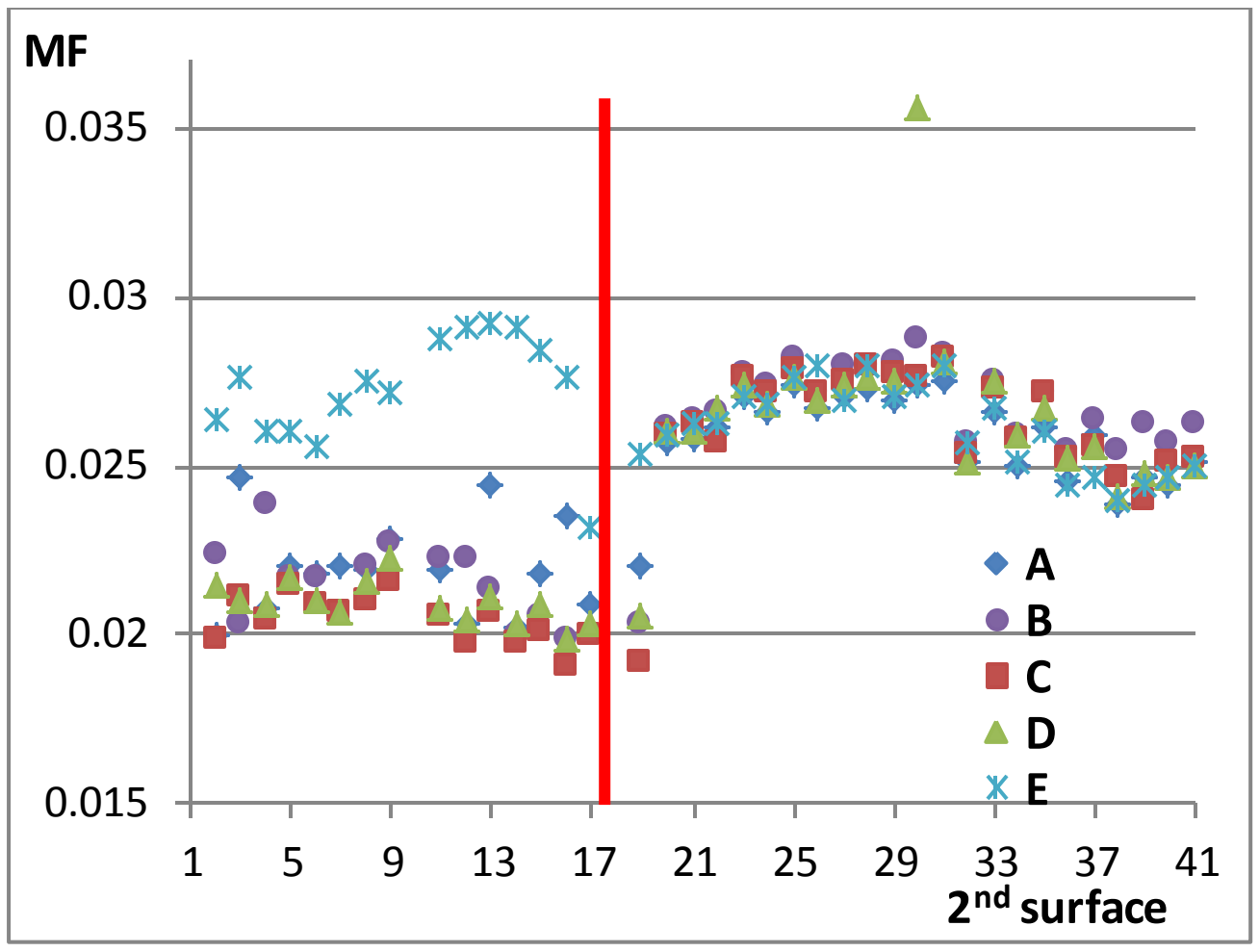

Fig. 11 Optimization strategy for the lithographic system. The first freeform surface is surface 18, and is marked in the diagram with a red line. The y axis is the value of the merit function, and the x axis represents the surface number of the second freeform surface. Results obtained from different strategies are marked with different signs.

**B. Two freeform mapping**

As the results above suggest, optimizing freeforms simultaneously or successively with fixed orders do not affect the final performance very much. It makes sense in the optical design process to add the freeform surfaces separately in steps. To be noted, the condition of system is changed after inserting the first freeform. As indicated in Fig. 12, the green curve shows the second freeform performance curve after making surface 3 the first freeform, while purple and red curves correspond to the cases where the first freeform is surface 16 or surface 20 respectively. As we can see, the curves are completely different. The positions of the second freeform close to the first freeform surface seem to be worse choices opposed to those away from it. The change of the system conditions after the previous freeform insertion shows the necessity of running the macro again before the insertion of the next freeform. We also notice the new one-freeform curves are no longer similar to the eccentricity curve anymore due to the change of system conditions.

We run all possible two-freeform-surface combinations for the Scheimpflug system, and optimize them as freeforms successively each for 50 cycles. The corresponding merit function values that represent the overall system performance for each surface combination are plotted in Fig. 13. In this figure, the y axis is the number of the surface that is made freeform first, while the x axis is the number of the surface that is made freeform second. The color of each small square in the picture corresponds to the system merit function value after optimization with the corresponding two-freeform combination in the specified sequence. The white stripes in the diagram are zero positions that correspond to stop position surface 6 and positions where the first and the second surfaces coincident.

To verify the influence of freeform sequence, we use the value of Fig. 13 to minus its upper half and get Fig. 14. In Fig. 14, the lower half corresponds to the differences caused by different optimization sequence of the same freeform combination. Compared with the system performance in Fig. 13, the differences are relatively small in scale. When the first freeform is chosen around stop position 6, the system performance is significantly worse with the second freeform set around surface 14. The explanation for that is straightforward. When we look at the best position for placing the first freeform for the Scheimpflug system in Fig. 6, surfaces around the stop are not preferred. When these surfaces are selected as the first freeform, it is more likely to trap the optimization process in a worse local minima than replacing a better surface location as the first freeform. This difference is especially obvious when locations of the second freeform surface happen to be the optimum options for inserting the first freeform. Similar phenomena is also seen when the second freeform is selected around the stop position. Therefore it makes sense to place the freeform surface at the optimal positions first, before placing the second freeform at other positions.

From Fig. 13 we can see that the best two-freeform combinations are surface 14 to surface 16 plus surface 1, which are basically the combination of surfaces from two minima of the curve in Fig. 6. In contrary, when two freeform locations are selected close to each other, which corresponds to the area around the white diagonal, the system performance is worse. This, together with the findings from Fig. 12, verifies that there are indeed interactions between freeform surfaces, when multiple freeforms exist in the same system. The best positions to place two freeforms are the best one-freeform position plus the new best one-freeform position after re-evaluation of the system.

We again apply the method on the lithographic system, the mapping of the two-freeform combinations is given in Fig. 16. The best two-freeform combination is again the best one-freeform position surface 18 plus the new best one-freeform position surface 17 shown in Fig. 15. Coincidently, the best locations are the only two mirror surfaces next to each other. This seems to violate the rule we just found about not putting two freeforms back to back. However, if we look closely at the system structure again, these two mirrors surfaces are next to the intermediate image planes that sit directly in between. Ray bundles from different fields can be separated quite well on the mirrors. Moreover, these two mirrors directly violate the system's rotational symmetry, which intuitively makes them better candidates for placing asymmetric freeform surfaces.

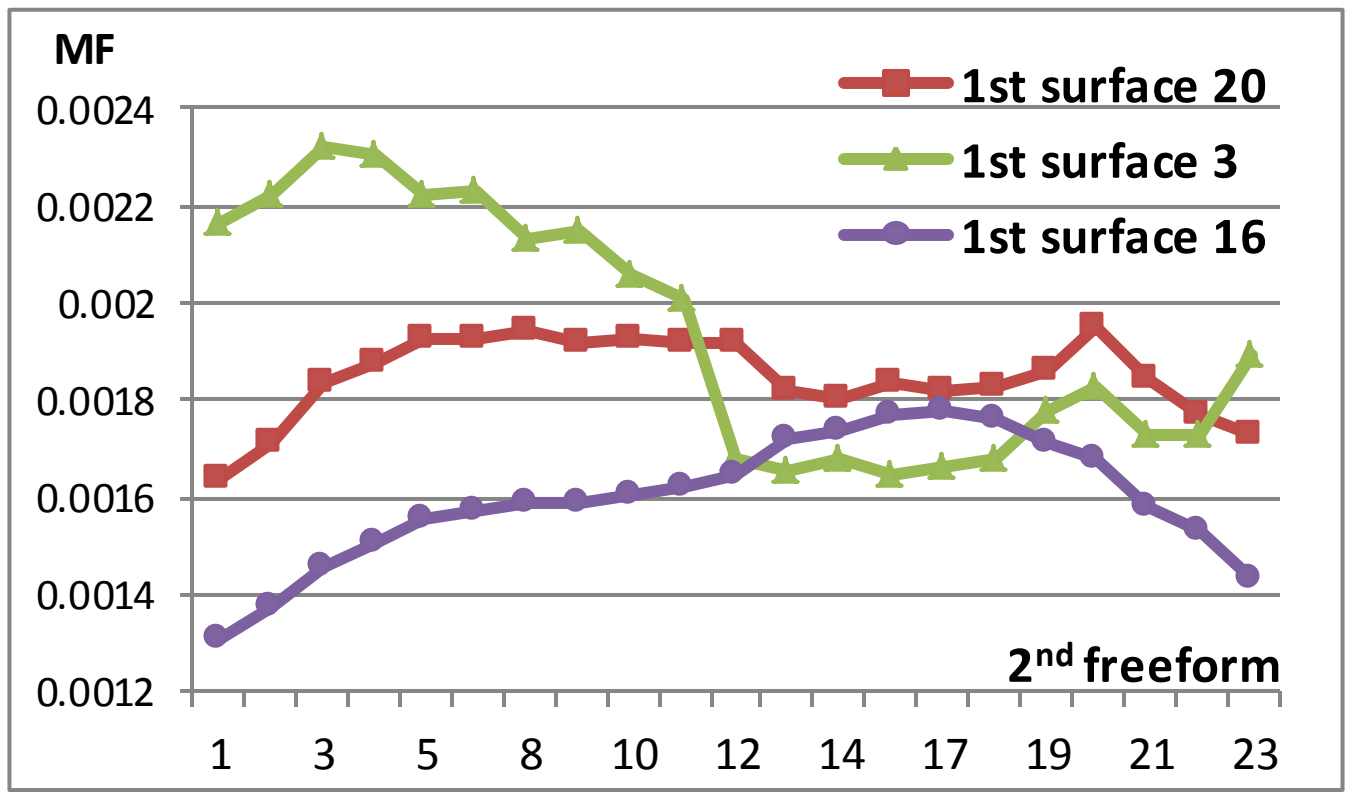

Fig. 12 Selection of the second freeform surface for the Scheimpflug system. The best location changes after placing the first freeform. When the first freeform surface is surface 16 with $\chi$ =-0.4, the second best freeform surface position is surface 1 with $\chi$ =-0.7.

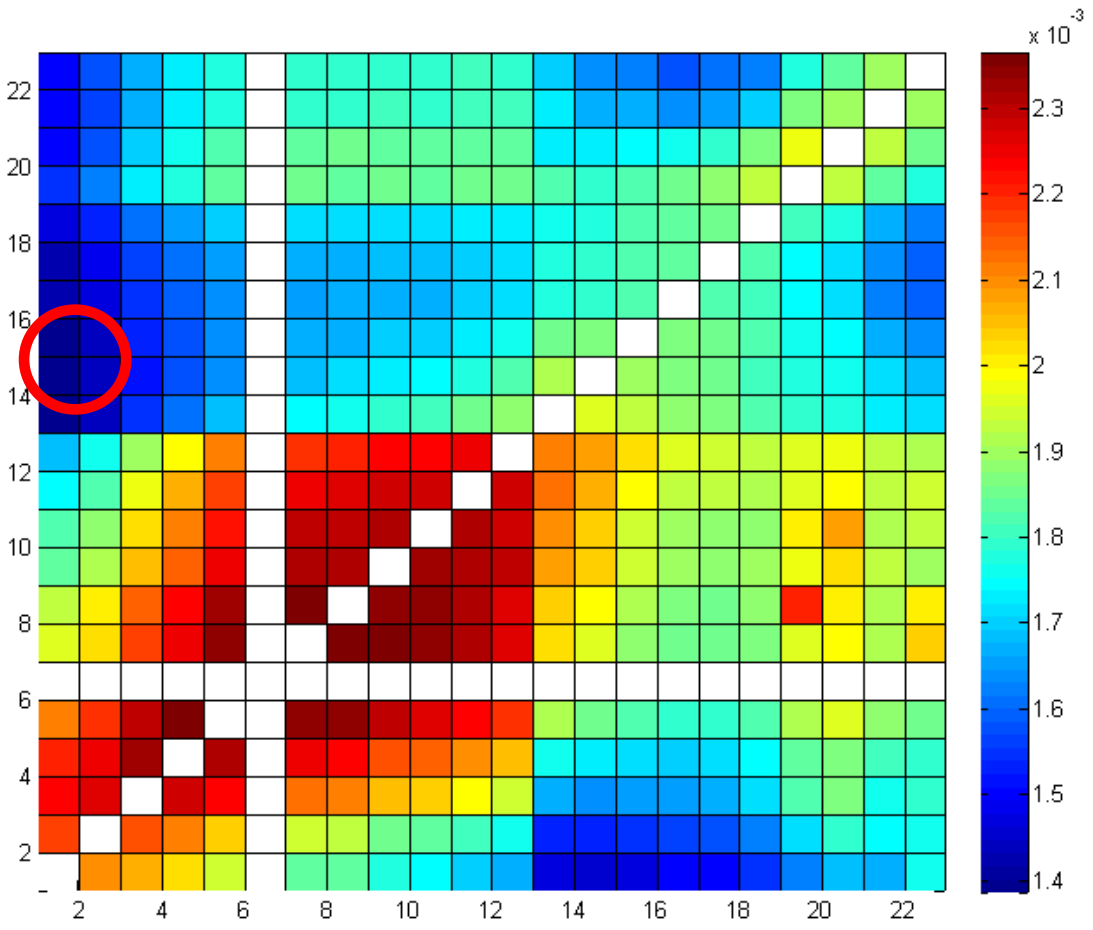

Fig. 13 Scheimpflug system, two freeform mapping. The y axis corresponds to the first freeform surface number, and the x axis corresponds to the second freeform surface number. The white strips in the figure are stop positions and positions where the first freeform is the same as the second freeform. The optimal two-freeform combination area is marked with a red circle.

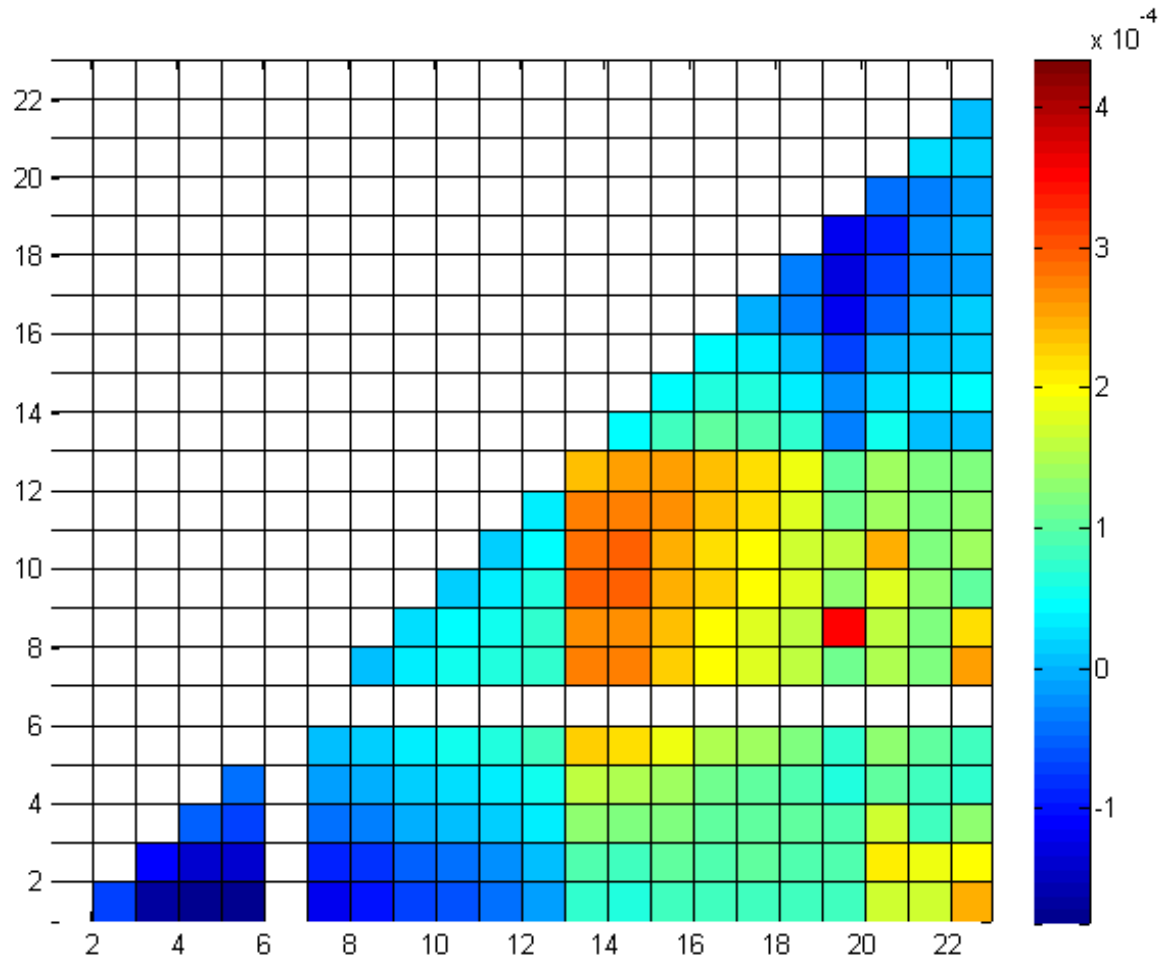

Fig. 14 Influence of freeform sequence for a Scheimpflug system. The lower half equals to the same part of Fig. 13 minus the upper half. The upper part equals to zero after the minus calculation.

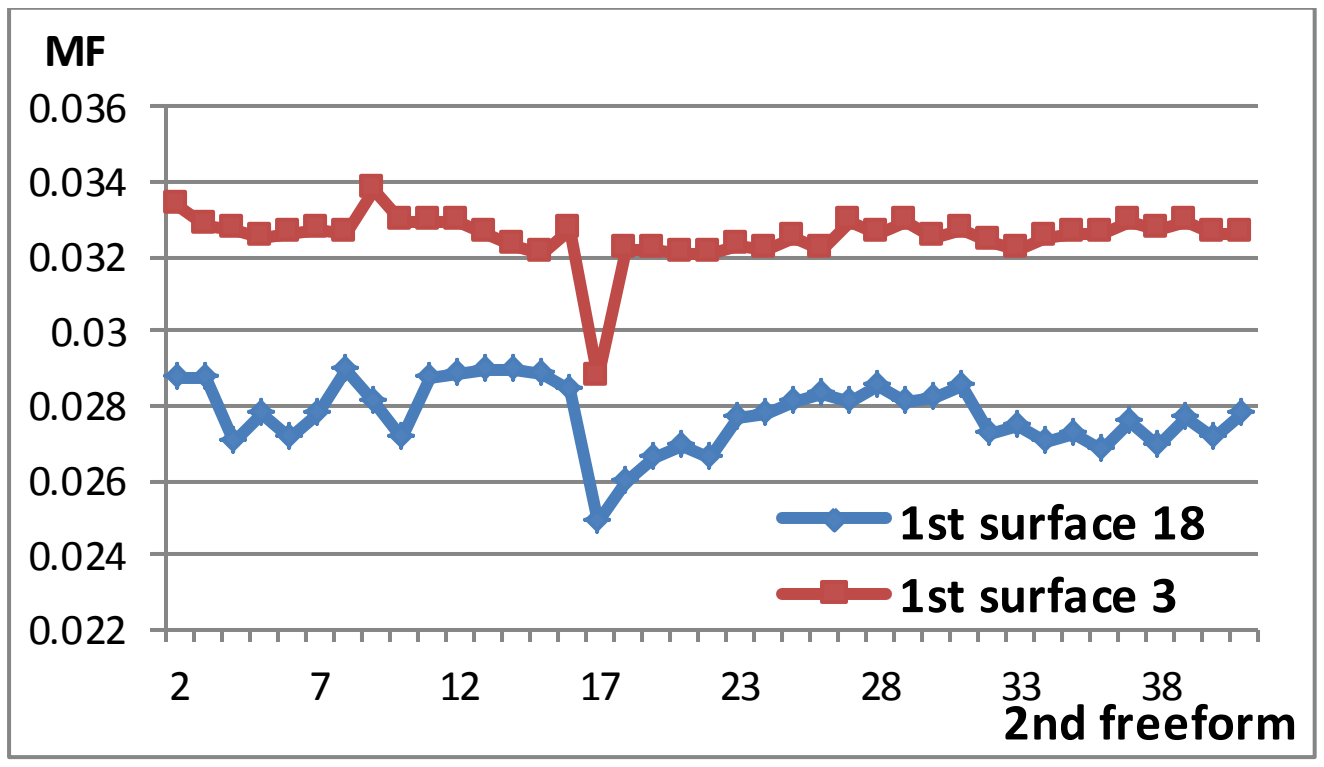

Fig. 15 Selection of the position of the second freeform surface for the lithographic system. When the first surface is selected as surface 18, the new best location for the next freeform surface is surface 17.

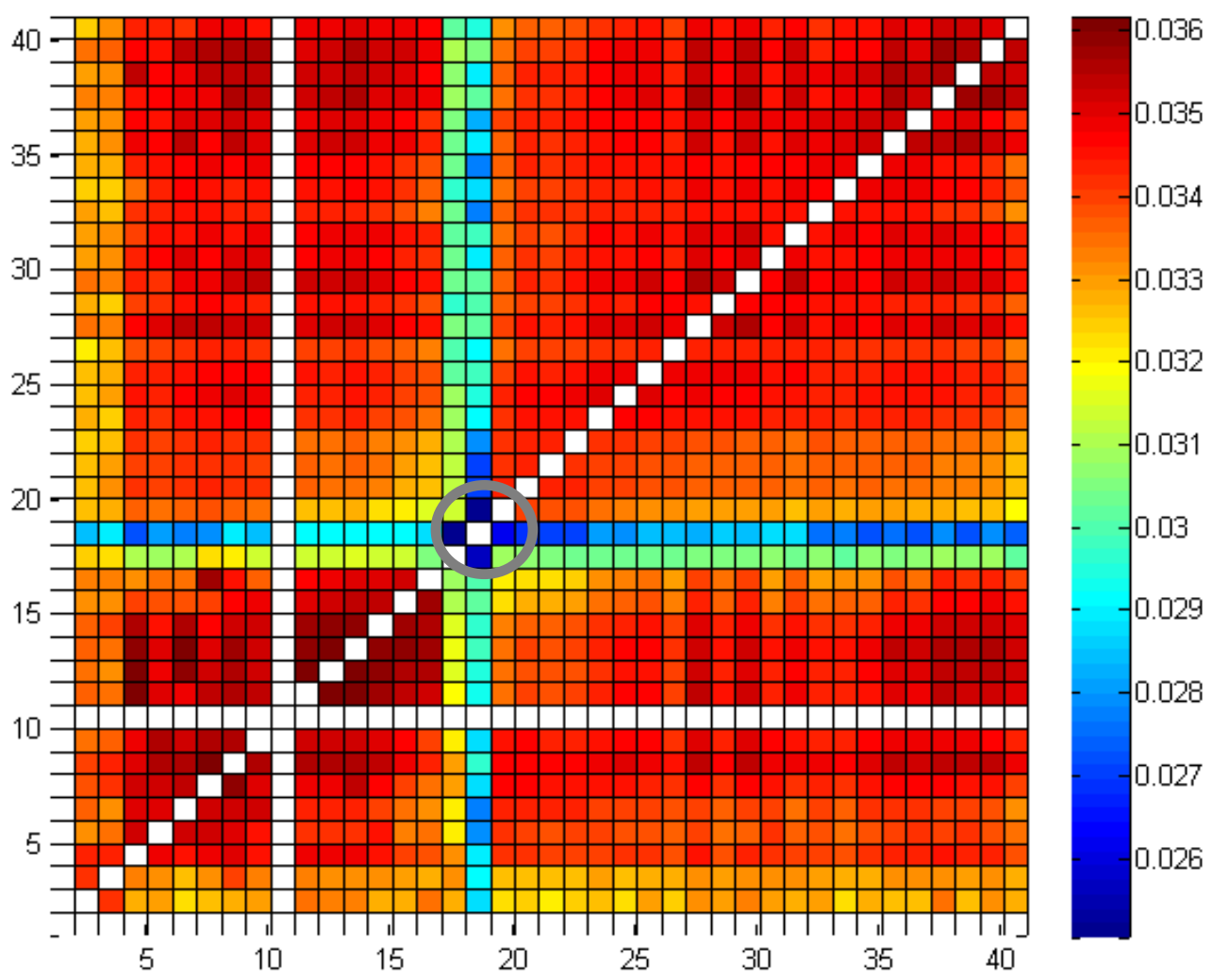

Fig. 16 Lithographic system, two freeform mapping. The y axis corresponds to the first freeform surface number, and the x axis corresponds to the second freeform surface number. The white strips in the figure are invalid positions. The optimal two-freeform combination area is marked with a grey circle.

## 5. Recommended work flow

To run all the possible surface combinations to determine the best locations is very time consuming. In the case of the lithographic system, it takes one day to generate the graph in Fig. 16. However, on the other hand it is much more convenient to generate the one-freeform position curve with appropriate variables. The agreement of results coming from two different methods, two-freeform mapping and one-freeform curve, has provided us a faster but still trustworthy way to locate the optimal freeform positions.

Summarizing our findings in the previous sections, we propose a workflow that can be efficiently used to replace system surfaces with freeforms. The workflow can thereby be applied directly without the requirement of special knowledge regarding the aberration theory for freeform surfaces. In the workflow given in Fig. 17, multiple freeform surfaces are added one after another until the system specifications are reached. Initial freeform orders are decided based on the system aberration constitution. While deciding the position to place the new freeform, only variables of the new freeform as well as the defocus of the system are used for the optimization. Not only the speed of the macro can be guaranteed with the minimum degrees of the solution space, but the rules for the optimal freeform position can also be seen more clear with respect to factors like stop position, previous freeform position and system layout.

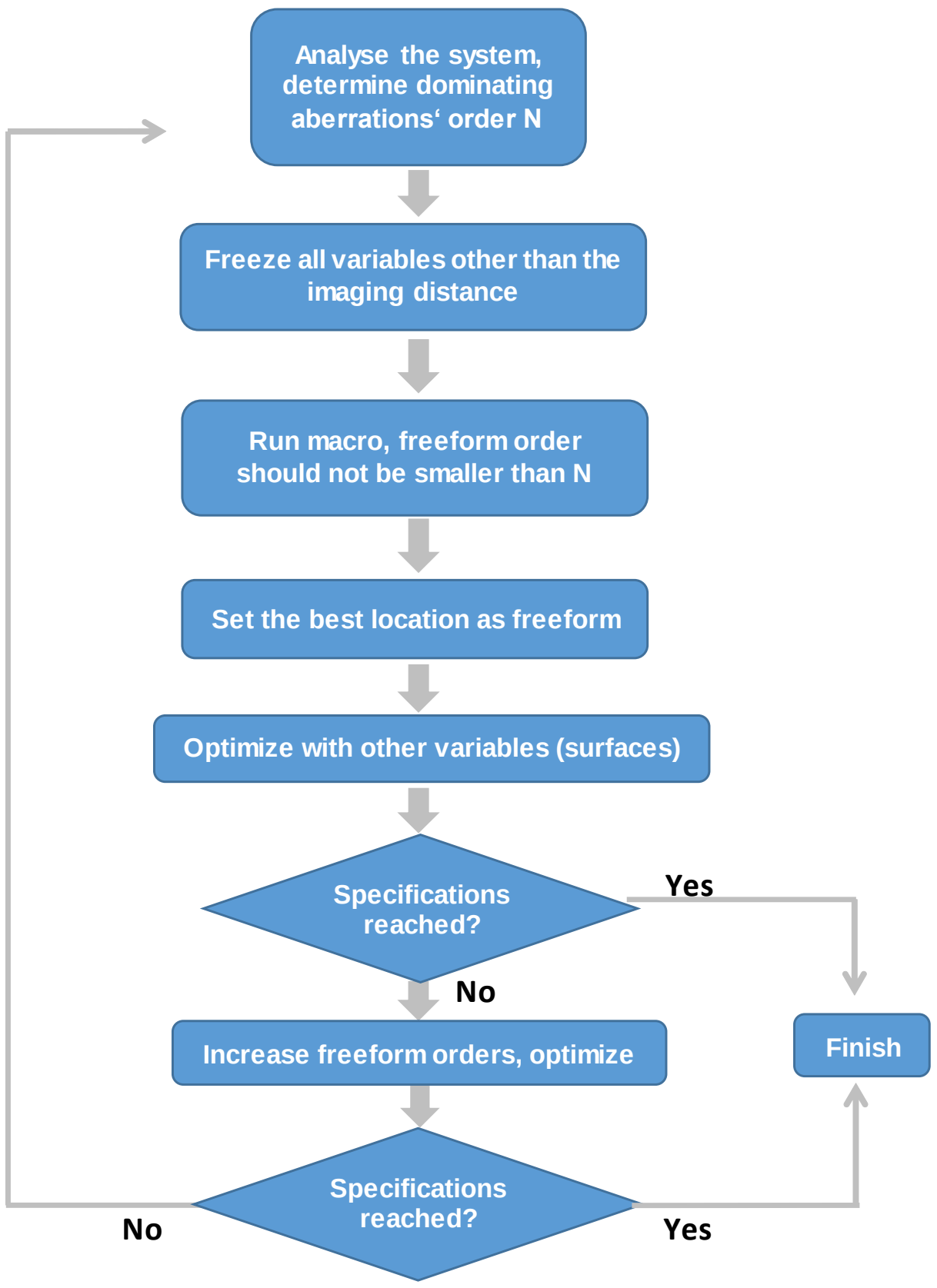

Fig. 17 Workflow for implementing freeform surfaces in a system.

Up until now, we have only studied the fast assessment method to determine the freeform surface locations using limited variables. To validate the workflow, we apply it on the Scheimpflug system to simulate the real design process. The system has already been optimized as good as possible with spherical surfaces before considering using freeforms. We follow the steps given in the workflow in Fig. 17 and replace two surfaces, surface 16 and surface 1 one after another to achieve a nearly diffraction limited performance as indicated in Fig. 18. The whole optimization process took us less than ten minutes.

We also show the full y field spot performance of the Scheimpflug system with respect to different freeform surface position selections in Fig. 19. As can be seen, the overall spot size of our freeform selection is not only significantly decreased compared with the original field performance, but also has an improved homogeneity. This improvement is further shown in Table **1**, where the necessity of the second freeform is seen directly. Comparing the field performance with some other freeform selections, the freeform surface positions selected according to our workflow seem to be the optimal solution for the design problem. Hence, the proposed workflow leads to a fast selection of the freeform position.

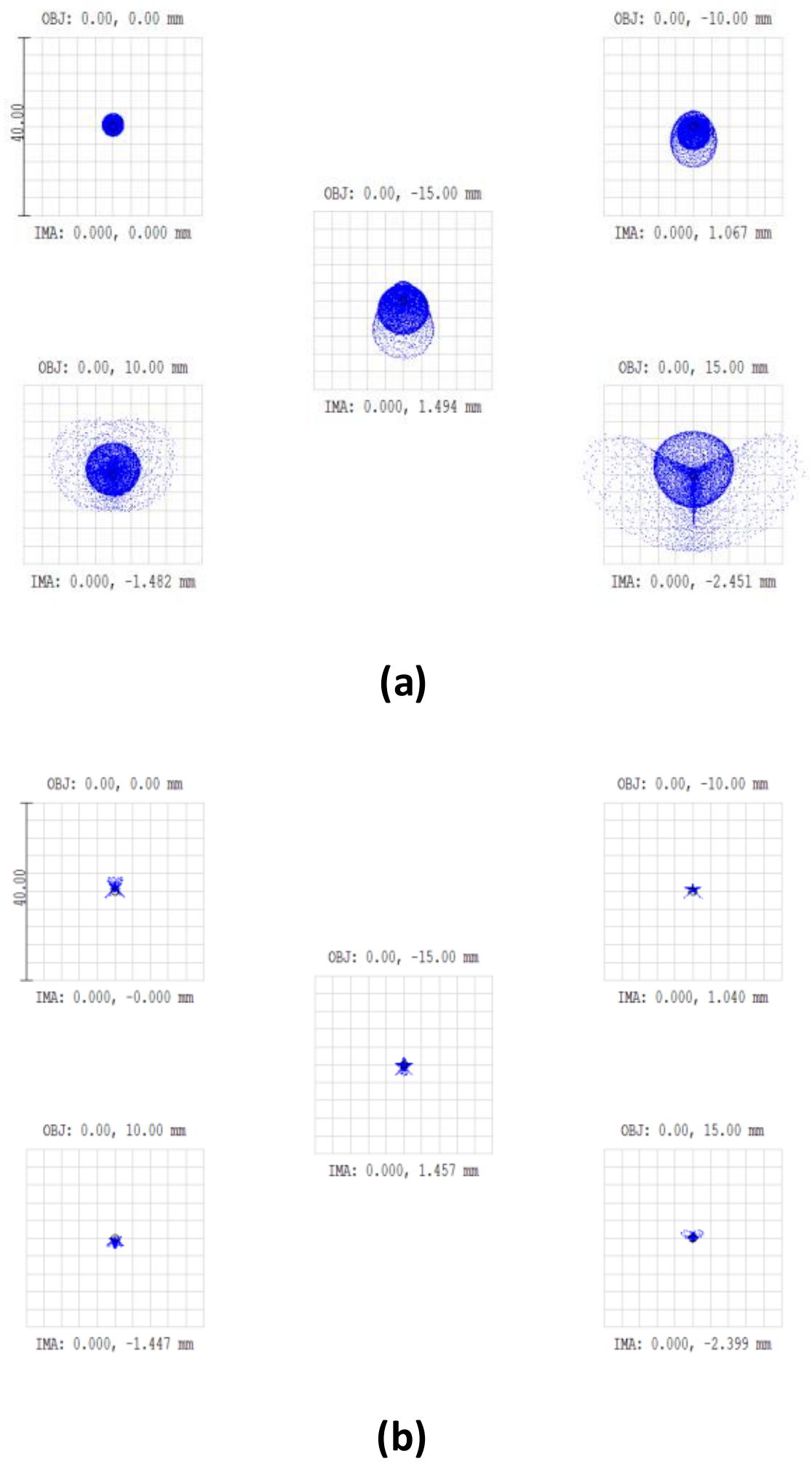

Fig. 18 Spot diagrams of the Scheimpflug system before (a) and after (b) optimization according to the workflow. The two diagrams have the same reference box size. The original system (a) has an averaged spot radius of 4.45 µm for the given fields, while the optimized system (b) has an averaged spot radius of 0.78 µm. The airy radius is 0.8 µm for (b), therefore the optimized system has a nearly diffraction limited performance.

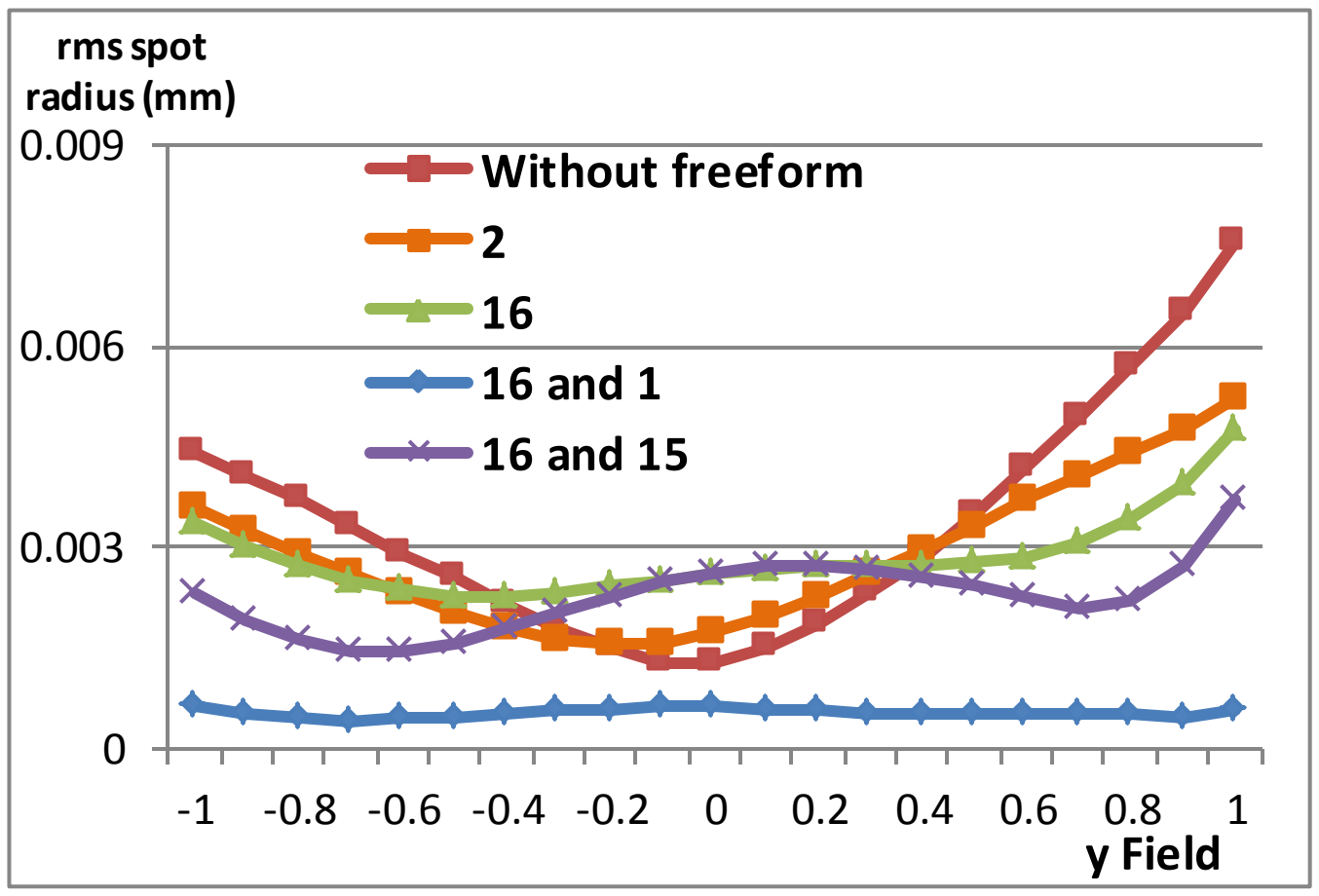

Fig. 19 Different field performance of the Scheimpflug system due to different freeform selections. The red curve is the field performance of the system with only spherical surfaces. With the best two-freeform selection surface 16 and surface 1 chosen, the overall spot size is not only decreased for the whole field, but also has an improved homogeneity.

Table 1 Scheimpflug system resolution improvement with freeforms. Both the averaged rms spot radius and the field uniformity are improved with one freeform surface (surface 16) and two freeform surfaces (surface 16 and 1).

| | **Original** | **1 freeform** | **2 freeform** |
|---|---|---|---|
| **Averaged rms radius** $\emptyset_{ave}$ | 3.33 µm | 2.86 µm | 0.53 µm |
| **Non-uniformity** $\emptyset_{\max}/\emptyset_{\min}$ | 5.84 | 2.07 | 1.48 |

## 7. Conclusions

In the content of this paper, we try to investigate general methods for multiple freeform optimization for complex multi-lens imaging systems. Two challenging systems with large field aberration contribution are investigated as examples.

We discuss the possible influences of freeform normalization radius, freeform order and eccentricity, and propose a method to determine the best location to place one freeform surface. Based on the findings of one-freeform location, we analyze and evaluate different strategies for implementing two freeform surfaces in complex optical systems. Fig. 10 and Fig. 11 show the impact of different optimization strategies. Both figures provide evidence that it is reasonable for optical designers to implement freeform surfaces successively. Moreover, the agreement between results coming from two-freeform mapping and one-freeform curves give us a fast and efficient way to assess the possible freeform locations. With the summarized workflow, we are able to optimize the Scheimpflug system faster and more efficiently than before to a nearly diffraction limited performance. This method, though being demonstrated to a limited number of systems, should have a much wider application on other complex imaging systems.